\documentclass[10pt,prd,singlecolumn,
nofootinbib, noshowkeys, noshowpacs, superscriptaddress, floatfix, aps]{revtex4-2}
\usepackage[dvips]{graphics,graphicx}
\usepackage[usenames,dvipsnames]{xcolor}
\definecolor{darkblue}{RGB}{0,0,196}
\definecolor{darkgreen}{RGB}{0,120,0}
\usepackage[colorlinks=true,linktocpage=true,linkcolor=darkblue,citecolor=red,urlcolor=darkblue]{hyperref}
\usepackage{framed}
\usepackage{mathtools}
\usepackage{float}
\usepackage{color}
\usepackage{slashed}
\usepackage{wrapfig}
\usepackage{tikz-feynman}
\usepackage{subcaption}
\usepackage{amsmath}
\usepackage{amssymb}
\allowdisplaybreaks
\usepackage{listings}
\usepackage{tikz-feynhand}
\usepackage{bm}
\usepackage{fixmath} 
\usepackage{braket}
\usepackage{multirow}
\usepackage{lipsum} 
\usepackage[toc,page]{appendix}
\usepackage{setspace}
\usepackage{simpler-wick}
\usepackage{stackengine}
\usepackage{bbm}
\usetikzlibrary{decorations.markings}

\newcommand{\no}{\nonumber}
\newcommand\hstar[1]{\ThisStyle{\ensurestackMath{%
			\setbox0=\hbox{$\SavedStyle#1$}%
			\stackengine{0pt}{\copy0}{\kern.2\ht0\smash{\SavedStyle\star}}{O}{c}{F}{T}{S}}}}

\newcommand{\tr}{\operatorname{Tr}}
\usepackage[normalem]{ulem}
\newif\ifshowedits
\showeditstrue
\newcommand{\Rev}[1]{\ifshowedits{\color{blue}#1}\else#1\fi}

\newcommand{\BlueEdit}[1]{\ifshowedits{\color{blue}#1}\else#1\fi}
\newcommand{\RedCut}[1]{\ifshowedits{\color{red}\sout{#1}}\fi}
\newcommand{\RedMark}[1]{\ifshowedits{\color{red}#1}\fi}
\newcommand{\RedMathCut}[1]{\ifshowedits\textcolor{red}{\ensuremath{#1}}\fi}
\begin{document}
	\title{ Thermal Fermion Propagators and Diagonal HTL Effective Action in the Early Universe} 
	\author{Ali Hataei\footnote{a.hataei@alumni.sbu.ac.ir}
		\affiliation{Department of Physics, Shahid Beheshti University,\\Tehran, Iran}}
	\author{S. S. Gousheh\footnote{ss-gousheh@sbu.ac.ir}}
	\affiliation{Department of Physics, Shahid Beheshti University,\\Tehran, Iran}
	\begin{abstract}
Within the minimally extended Standard Model (MESM), with Dirac neutrinos, we study fermion propagation in the electroweak-symmetric plasma, where the
		Higgs expectation value vanishes. We show that the zero-temperature mass
		eigenstates do not coincide with the left-handed fermionic quasiparticles
		identified by the thermal medium, while the right-handed fermions remain
		diagonal in this basis. At the leading hard-thermal-loop (HTL) order, the
		left-handed fermion self-energies form flavor matrices containing the 
	CKM and PMNS mixing structures.
	Diagonalizing flavor matrices, we show that the two members of each left-handed weak doublet possess identical leading-HTL thermal spectra. The eigenvalues are CP-even and are invariant under weak-basis transformations.
		In this symmetric-phase thermal basis, the leading HTL fermion propagators and
		electroweak gauge interactions are diagonal, while the flavor structure is
		transferred to the Yukawa interactions. We further express the HTL effective
		Lagrangian in this basis, derive the corresponding dispersion
		relations, and numerically compare the eigenvalues of the complete leading-HTL matrices with the diagonal
		thermal coefficients obtained in the broken-phase mass basis.
		This construction provides a diagonal HTL description of fermionic
		quasiparticles and a systematic framework for resummation and higher-order
		corrections in the electroweak-symmetric plasma.
	\end{abstract}
	\maketitle
	
			
	\section{Introduction}\label{sec. 1}
	Thermal field theory provides the framework for describing
	quantum fields in a medium
	\cite{Kapusta:2006pm,Bellac:2011kqa,Laine:2016hma,Mustafa:2022got,LANDSMAN1987141}, in which interactions with the surrounding medium dress the elementary fields and
	modify their propagation and effective interactions.
	At the weak coupling limit, with the temperature $T$ being the dominant
	scale, momenta separate into hard momenta of order $T$, soft momenta
	of order $\xi T$, and semihard momenta
	\cite{Wang_2000,mzw1-bsvy,PhysRevLett.134.011905}
	satisfying $T\gg Q\gg \xi T$, where $\xi$ denotes a generic weak
	coupling \cite{Hataei}.
	The medium is dominated by hard excitations, while soft modes at the
	collective scale probe the response of these hard thermal constituents.
	The resulting medium effects are central to transport and damping
	\cite{PhysRevD.109.096011,PhysRevD.47.5589,Peign__1993,PhysRevD.51.862,PhysRevLett.64.1338},
	photon and dilepton production \cite{PhysRevD.38.2814,Arnold_2001},
	heavy-quark dynamics and parton energy loss
	\cite{Mustafa_2005,Mustafa_2006,Chakraborty_2006,Mustafa:2022got},
	early-Universe physics
	\cite{LEUTWYLER19911,PhysRevLett.44.631,HAWKING198235,Matsubara:1955ws,
		Takahashi:1996zn,Davidson_1994,RevModPhys.93.035004,PhysRevD.107.035021,
		bernal2024thermaldarkmatterlowtemperature,Katz_2014,BALDES2018373,
		arcadi20232hdamodelcolliderdark}, high-energy particle collisions
	\cite{PhysRevD.47.5138}, neutron-star matter \cite{Kapusta:2006pm},
	and thermodynamics \cite{Mustafa:2022got}.

	A distinctive feature of the collective scale is that the leading thermal
	self-energies and vertices become parametrically as important as the
	corresponding bare inverse propagators and vertices. Higher-loop diagrams can consequently contribute at the same parametric order as lower-loop diagrams. The Hard Thermal Loop (HTL) resummation framework provides a
	systematic reorganization of perturbation theory in which contributions of the same parametric order are incorporated consistently. As an example, diagrams containing
	one soft loop and an arbitrary number $N$ of leading hard thermal subgraphs
	contribute at the same order and are collectively represented by the
	corresponding one-soft-loop diagram constructed from HTL-resummed propagators
	and vertices \Rev{\cite{BraatenPisarski:1990soft}}. This resummed description is essential for obtaining consistent
	gauge-independent results and for the cancellation of divergences among
	contributions of the same parametric order; see Ref.~\cite{Hataei} for a
	detailed discussion.
	The gauge-invariant HTL effective
	Lagrangian was developed to generate the leading thermal self-energies together with the higher-point effective vertices
	\cite{PhysRevD.45.R1827,BRAATEN1990310,Taylor:1990ia,FRENKEL1992156,Mustafa:2022got,PhysRevLett.83.2139,PhysRevD.66.085016,Andersen_2011,Haque_2014,PhysRevD.89.061701}.

	For theories with several fermion generations, the construction of the
	HTL effective theory also raises a question of basis.
	In the broken electroweak phase, where
	$\mathrm{SU}(2)_L\times\mathrm{U}(1)_Y$ is reduced to
	$\mathrm{U}(1)_{\rm EM}$ through the Higgs mechanism
	\cite{HIGGS1964132,PhysRev.145.1156,PhysRevLett.13.508}, the familiar
broken-phase mass basis provides a natural description of the
	propagating fermions.
		At temperatures above the electroweak transition,
	$T>T_{\mbox{\tiny EWPT}}$, the symmetry is restored
	\cite{Kirzhnits:1972ut,PhysRevD.9.3320,PhysRevD.9.3357,
		Kirzhnits:1976ts,AD-Linde_1979,DOnofrioRummukainen:2016}, as thermal corrections render the Higgs
	mass-squared positive \cite{PhysRevD.45.2933}. The Higgs expectation value
	then vanishes, while all four real degrees of freedom contained in
	$\Phi=\mathrm{Col}(\phi^{(+)},\phi^{(0)})$ remain active in the plasma.
	The Yukawa interactions nevertheless contribute to the thermal
	self-energies. Consequently, the broken-phase mass basis need not
	diagonalize the leading thermal mass matrices, and the corresponding
	HTL-resummed propagators become matrix valued in generation space. This motivates identifying the symmetric-phase thermal basis selected by the thermal
	medium.

	In the symmetric phase, earlier studies adopted the zero-temperature mass
	basis and demonstrated the appearance of off-diagonal flavor structures in
	the thermal fermion self-energies, although these contributions were treated
	perturbatively or neglected in the analysis of the dispersion relations
	\cite{Gavela_1994,Quimbay:1995}. Our contribution is to construct explicitly the common thermal propagation
	basis of the weak doublets and the resulting mode-diagonal HTL fermionic
	effective action, while retaining the transformed Yukawa flavor structure.
The recently developed AutoTherm framework automates HTL-resummed thermal production rates from \(2\leftrightarrow2\) processes, with applications to gravitino and axino production in the early Universe~\cite{AutoTherm2026,AutoThermGravitino2026}. Its current implementation requires diagonal fermionic thermal-mass matrices in generation space.
	
	In this work, we first represent the MESM in the broken-phase mass basis,
	where the Cabibbo--Kobayashi--Maskawa (CKM) \cite{Cabibbo:1963,KobayashiMaskawa:1973} and Pontecorvo--Maki--Nakagawa--Sakata (PMNS) \cite{Pontecorvo:1957,MakiNakagawaSakata:1962} matrices appear not only in the weak interactions but
	also in the Yukawa sector. In this basis, the Yukawa channels generate
	Hermitian but non-diagonal thermal matrices, whose diagonal entries therefore
	do not represent the thermal masses of independent
	quasiparticles. At the leading HTL order, the momentum dependence of the fermion
	self-energy factorizes from its generation-space matrix, allowing the complete
	propagation problem to be diagonalized by a single momentum-independent
	unitary transformation. This defines the symmetric-phase thermal basis. We further
	show that the thermal matrices of the two members of each left-handed weak
	doublet are unitarily  equivalent and hence have identical thermal spectra.
	The leading resummed propagators and electroweak gauge interactions are
	therefore diagonal in the symmetric-phase thermal basis, while the mixing structure is
	transferred entirely to the transformed Yukawa vertices. The fermionic HTL
	effective action can consequently be organized mode by mode, with the usual
	HTL momentum dependence and Ward identities applying independently to each
	thermal eigenmode. No additional rotation is introduced for the right-handed
	singlets: their leading HTL thermal matrices are already diagonal in the
	original broken-phase mass-basis convention.
	We derive the thermal eigenvalues analytically from basis-independent
	invariants, construct the corresponding diagonal HTL effective Lagrangian,
	propagators, gauge vertices, and dispersion relations, and numerically
	compare the exact thermal eigenvalues with the diagonal entries obtained
	in the broken-phase mass basis. A central point is that the thermal
	spectrum is rephasing invariant and also CP-even, while the corresponding thermal
	eigenvectors retain the expected basis dependence.
	
	The paper is organized as follows. In \BlueEdit{Sec.}~\ref{sec. 2}  we write the MESM
	Lagrangian in the electroweak-symmetric phase in the broken-phase mass basis.  In Sec.~\ref{sec. 3} we derive the corresponding one-loop HTL
	self-energies and thermal mass matrices.  In Sec.~\ref{sec.4} we construct the
	symmetric-phase thermal basis. Section \ref{sec.conclusion} concludes the main results. 
	\section{The Minimally Extended Standard Model in the Symmetric Phase}\label{sec. 2}
	In the MESM, the asymmetry between the lepton and quark sectors in the SM, due to the absence of right-handed neutrino fields, is removed \cite{Giunti:2007ry}. We assume three Dirac neutrinos and set gauge-invariant right-handed Majorana mass terms to zero. For clarity we suppress the QCD Lagrangian in this section, but restore its leading thermal self-energy contribution in Sec.~\ref{sec. 3}. We write the MESM Lagrangian, with the effective Higgs mass, as
	\begin{align}
		\mathcal{L}
		=\mathcal{L}_{Q}+\mathcal{L}_L+\mathcal{L}_{\mathrm{gauge}}
		+\Bigl\{(D_{\mu}\Phi)^\dagger D^\mu\Phi
		-\mu^2(T)\Phi^\dagger\Phi
		-\lambda(\Phi^\dagger\Phi)^2\Bigr\},
	\end{align}
	where $\mathcal{L}_Q$, $\mathcal{L}_L$, and $\mathcal{L}_{\mathrm{gauge}}$ denote the quark, lepton, and gauge-field parts of the MESM Lagrangian introduced below, respectively.
	Here, $D_{\mu}=\partial_\mu+ig A^a_\mu \tau_a+ig^\prime \frac{Y}{2}B^Y_\mu$ is the covariant derivative, where $\tau_a=\sigma_a/2$ $(a=1,2,3)$ are the $\mathrm{SU}(2)_L$ generators, $A^a_\mu$ are the corresponding gauge fields, and $B^Y_\mu$ is the $\mathrm{U}(1)_Y$ gauge field. 
	The curly bracket contains the free and self-interacting terms of the Higgs doublet, including the effective mass-squared parameter $\mu^2(T)$, which is positive in the symmetric phase\footnote{At zero temperature and tree level, the Higgs mass-squared at $\phi=0$ is negative, leading to spontaneous symmetry breaking and a nonzero vacuum expectation value of $\phi$. However, the leading thermal correction introduces a positive $T^2$-dependent contribution, which selects the symmetric-phase minimum at sufficiently high temperature (a smooth crossover for physical Standard Model parameters) \cite{PhysRevD.45.2933, PhysRevD.50.6662}.}.
	The coupling constant $\lambda$ also has thermal corrections \cite{Anderson:1991zb, PhysRevLett.69.1304}, whose effect is negligible for the present analysis.		
	The Higgs doublet is defined as 
		$\Phi=(\phi^{(+)},\phi^{(0)})^{\mathrm{T}}$,
	 where $\phi^{(+)}$ and $\phi^{(0)}$ are the charged and neutral complex scalar fields.
	The chiral fermion fields for the three generations are 
	\begin{equation}\label{2.2}
		\begin{split}
			&Q_i\equiv\begin{pmatrix}
				u_{i_L}\\
				d_{i_L}\\
			\end{pmatrix},\quad
			L_i\equiv\begin{pmatrix}
				\nu_{i_L}\\
				e_{i_L}\\
			\end{pmatrix},\quad u_{i_R},\quad d_{i_R},\quad \nu_{i_{R}},\quad	e_{i_R},\quad i=1,2,3,
		\end{split}
	\end{equation}
	where $Q_i$ and $L_i$ are the left-handed quark and lepton doublets, while $u_{i_R}$, $d_{i_R}$, $\nu_{i_R}$, and $e_{i_R}$ are the corresponding right-handed singlets. In what follows, we suppress the generation indices and use matrix notation for the fields.
	
The broken-phase mass basis, in which the Yukawa matrices are diagonal, is obtained from a generic weak-interaction basis by unitary chiral rotations of the left- and right-handed fermion fields as:
	\begin{align}
		&Q\rightarrow U_QQ^\prime
		=\mathrm{diag}(U_u,U_d)\,Q^\prime,\qquad
		u_R\rightarrow W_u u_R^\prime,\qquad
		d_R\rightarrow W_d d_R^\prime,\no\\
		&L\rightarrow U_LL^\prime
		=\mathrm{diag}(U_\nu,U_e)\,L^\prime,\qquad
		\nu_R\rightarrow W_\nu\nu_R^\prime,\qquad
		e_R\rightarrow W_e e_R^\prime.
	\end{align}
	We denote the Yukawa coupling matrices in the weak basis by $\Lambda_d$, $\Lambda_u$, $\Lambda_e$, and $\Lambda_\nu$ for down-type quarks, up-type quarks, charged leptons, and neutrinos, respectively. The corresponding diagonal couplings in the broken-phase mass basis are
	\begin{align}
		&\lambda_d=U_d^\dagger \Lambda_d W_d
		=\mathrm{diag}(\lambda_{d_1},\lambda_{d_2},\lambda_{d_3}),\qquad
		\lambda_u=U_u^\dagger \Lambda_u W_u
		=\mathrm{diag}(\lambda_{u_1},\lambda_{u_2},\lambda_{u_3})\no\\
		&\lambda_\nu=U_\nu^\dagger\Lambda_\nu W_\nu
		=\mathrm{diag}(\lambda_{\nu_1},\lambda_{\nu_2},\lambda_{\nu_3}),\qquad
		\lambda_e=U_e^\dagger\Lambda_e W_e
		=\mathrm{diag}(\lambda_{e_1},\lambda_{e_2},\lambda_{e_3}),
	\end{align}
	where the chiral rotation matrices are unitary, $U^\dagger_{x}U_x=W^\dagger_x W_x=\mathbbm{1}$, with $x\in \{u,d,e,\nu\}$.
	The phases of the diagonal Yukawa eigenvalues are absorbed into the right-handed fields, so we take $\lambda_{x_i}\geq0$; consequently, $\lambda_x^\dagger\lambda_x=\lambda_x^2$ in the convention used below.
	Applying these transformations gives the quark-sector Lagrangian as
	\begin{align}\label{quark-sector-weak}
		\mathcal{L}^\prime_Q
		=i\bar{Q}^\prime(U_Q^\dagger\slashed{D}U_Q)Q^\prime
		+i\bar{u}_R^\prime(W_u^\dagger\slashed{D}W_u)u_R^\prime
		+i\bar{d}_R^\prime(W_d^\dagger\slashed{D}W_d)d_R^\prime 
		-\Bigl\{
		\bar{Q}^\prime(U_Q^\dagger\Lambda_d\Phi W_d)d_R^\prime
		+\bar{Q}^\prime(U_Q^\dagger\Lambda_u\tilde{\Phi}W_u)u_R^\prime
		\Bigr\}
		-\mathrm{H.C.},
	\end{align}
	where H.C. denotes the Hermitian conjugate of the terms in the curly bracket, and $\tilde{\Phi}:=i\sigma_2\Phi^*$.
	Since $W_u$ and $W_d$ commute with the relevant gauge generators, they drop out by the unitarity of the chiral transformations. As in the broken phase, the left-handed term gives
	\begin{equation*}
		\begin{split}
			i\bar{Q}^\prime(U_Q^\dagger\slashed{D}U_Q)Q^\prime
			&=i\bar{Q}^\prime
			\begin{pmatrix}
				\slashed{\partial}+\frac{ig}{2}\slashed{A}_3+\frac{i}{2}g^\prime\slashed{B}_Y Y
				&
				V\frac{ig}{2}(\slashed{A}_1-i\slashed{A}_2)
				\\
				V^\dagger\frac{ig}{2}(\slashed{A}_1+i\slashed{A}_2)
				&
				\slashed{\partial}-\frac{ig}{2}\slashed{A}_3+\frac{i}{2}g^\prime\slashed{B}_Y Y
			\end{pmatrix}
			Q^\prime .
		\end{split}
	\end{equation*}
	Here, $V=U_u^\dagger U_d$ is the CKM matrix \cite{Cabibbo:1963,KobayashiMaskawa:1973}, which contains one physical CP-violating phase for three generations \cite{PhysRevLett.53.1802}. 
	The first term in the curly bracket of Eq.~\eqref{quark-sector-weak} also becomes
	\begin{align*}
		U_Q^\dagger\Lambda_d\Phi W_d
		&=
		\begin{pmatrix}
			U_u^\dagger\Lambda_d W_d\,\phi^{(+)}
			\\
			U_d^\dagger\Lambda_d W_d\,\phi^{(0)}
		\end{pmatrix}
		=
		\begin{pmatrix}
			V&0\\
			0&\mathbbm{1}
		\end{pmatrix}
		\lambda_d
		\begin{pmatrix}
			\phi^{(+)}
			\\
			\phi^{(0)}
		\end{pmatrix}
		\equiv \mathcal{V}^\prime\lambda_d\Phi .
	\end{align*}
	The second term follows by the replacements
	\begin{align*}
		\mathcal{V}^\prime\rightarrow
		\mathcal{V}^{\prime\prime}
		=\mathrm{diag}(\mathbbm{1},V^\dagger),
		\qquad
		\lambda_d\rightarrow\lambda_u,
		\qquad
		\Phi\rightarrow\tilde{\Phi}.
	\end{align*}
	The same prescription applies to the lepton sector. The fermionic Lagrangian in the broken-phase mass basis is therefore
	\begin{subequations}\label{mass-basis-lagrangian}
		\begin{align}
			&\mathcal{L}^\prime_Q
			=i\bar{Q}^\prime(U_Q^\dagger\slashed{D}U_Q)Q^\prime
			+i\bar{u}^\prime_R\slashed{D}u^\prime_R
			+i\bar{d}^\prime_R\slashed{D}d^\prime_R 
			-\Bigl\{
			\bar{Q}^\prime\mathcal{V}^\prime\lambda_d\Phi d_R^\prime
			+\bar{Q}^\prime\mathcal{V}^{\prime\prime}\lambda_u\tilde{\Phi}u_R^\prime
			\Bigr\}
			-\mathrm{H.C.},\label{quark-sector-mass}\\
			&\mathcal{L}^\prime_L
			=i\bar{L}^\prime(U_L^\dagger\slashed{D}U_L)L^\prime
			+i\bar{e}^\prime_R\slashed{D}e^\prime_R
			+i\bar{\nu}^\prime_R\slashed{\partial}\nu^\prime_R 
			-\Bigl\{
			\bar{L}^\prime\mathcal{U}^\prime\lambda_e\Phi e_R^\prime
			+\bar{L}^\prime\mathcal{U}^{\prime\prime}\lambda_\nu\tilde{\Phi}\nu_R^\prime
			\Bigr\}
			-\mathrm{H.C.},
		\end{align}
	\end{subequations}
	with
	\begin{align*}
		\mathcal{U}^\prime=\mathrm{diag}(U,\mathbbm{1}),
		\qquad
		\mathcal{U}^{\prime\prime}=\mathrm{diag}(\mathbbm{1},U^\dagger).
	\end{align*} 
	Here, we define
	\[
	U\equiv U_\nu^\dagger U_e=U_{\rm PMNS}^\dagger,
	\]
	where $U_{\rm PMNS}=U_e^\dagger U_\nu$ denotes the conventional PMNS matrix. Thus, $U$ plays the same role in the lepton sector as
$V=U_u^\dagger U_d$ does in the quark sector.
	Equation~\eqref{mass-basis-lagrangian} shows that the flavor-mixing matrices enter both the charged-current and charged-scalar Yukawa interactions. In the broken-phase mass basis, the physical neutral-Higgs Yukawa couplings are flavor diagonal at tree level~\cite{Peskin:1995ev}; in unitary gauge, the Goldstone fields are absent. Their Yukawa-sector appearance originates from the charged component $\phi^{(+)}$. To see this explicitly, consider the first term in the curly bracket of Eq.~\eqref{quark-sector-mass},
	\begin{equation}\label{2.13}
		-\bar{u}^\prime_L V\lambda_d\phi^{(+)}d_R^\prime
		-\bar{d}^\prime_L\lambda_d\phi^{(0)}d_R^\prime .
	\end{equation}
The first interaction contains the CKM matrix and therefore mixes quark generations when $\phi^{(+)}$ is exchanged.    
	After removing $\nu_R$, the lepton-sector Lagrangian becomes
	\begin{equation}\label{2.19}
		\mathcal{L}^\prime_{L,\mathrm{SM}}
		=i\bar{L}^\prime\slashed{D}L^\prime
		+i\bar{e}_R^\prime\slashed{D}e_R^\prime
		-\bar{L}^\prime\lambda_e\Phi e_R^\prime
		-\bar{e}_R^\prime\Phi^\dagger\lambda_e L^\prime .
	\end{equation}
	Here, $\lambda_e$ is the diagonal charged-lepton Yukawa matrix. Thus, in the SM with massless neutrinos, the the weak interaction and mass bases can be aligned for the leptons.
	\section{The Fermion Propagators and Their Thermal Masses}\label{sec. 3}
	We assume a homogeneous, isotropic plasma in thermal equilibrium with vanishing chemical potentials and no flavor-dependent background potentials. We consider an arbitrary soft external momentum, $p^0,|\vec p|\ll T$. In the high-temperature regime considered here, the
temperature is the dominant scale, such that
$T\gg\{p^0,|\vec p|,\mu(T)\}$, where $\mu(T)$ is the effective Higgs mass.
In the electroweak-symmetric phase,
$\mu^2(T)\simeq c^2+\Pi_\Phi(0)$, where $c^2<0$ is the zero-temperature
tree-level Higgs mass-squared parameter and $\Pi_\Phi(P)$ denotes the
finite-temperature Higgs self-energy \cite{PhysRevD.45.2933}. The latter
receives gauge, scalar, and Yukawa contributions; in particular, the
dominant Yukawa contribution is that of the top quark,
$\Pi_\Phi^{(t)}(0)=\frac{1}{4}\lambda_{u_3}^2T^2$, with
$\lambda_{u_3}\simeq1$ \cite{PhysRevD.45.2933}. Although the individual thermal
contributions to $\Pi_\Phi(0)$ can be significant, the negative tree-level
parameter $c^2$ partially compensates the positive thermal correction.
Consequently, in the temperature range considered here, the resulting
effective mass $\mu(T)$ can remain small compared with the hard thermal
scale $T$.

The leading thermal contribution to the loop integrals arises from hard loop momenta, $k^0,|\vec k|\sim T$. Since both the external momentum and the
effective Higgs mass are soft compared with this hard scale, the HTL
expansion corresponds to
$|\vec k|\sim T\gg\{|\vec p|,\mu(T)\}$, and the dependence on $\mu(T)$
inside the hard loop contributes only beyond leading HTL order
\cite{Bellac:2011kqa,Carrington_1999}. The resulting leading HTL self-energy
of a left- or right-handed fermion takes the standard form
\cite{Bellac:2011kqa,PhysRevD.26.2789,PhysRevD.81.025014,Wang:2004tg}
\begin{equation}\label{3.3}
	\Sigma_{L,R}(P)
	=\frac{m^2_{f_{L,R}}(T)}{|\vec p|}
	\left\{
	\gamma_{0}\mathcal{Q}_0\left(\frac{p_0}{|\vec p|}\right)
	+\vec{\gamma}\cdot\hat{p}
	\left[
	1-\frac{p_0}{|\vec p|}
	\mathcal{Q}_0\left(\frac{p_0}{|\vec p|}\right)
	\right]
	\right\}\mathcal{P}_{L,R},
\end{equation}
where $p_0$ and $\vert\vec{p}\vert$ denote the energy and magnitude of the external three-momentum, respectively. The scalar function entering the retarded self-energy is
\begin{equation*}
	\mathcal{Q}_0\left(\frac{p_0}{\vert\vec{p}\vert}\right)
	=\frac{1}{2}\ln\left|\frac{p_0+\vert\vec{p}\vert}{p_0-\vert\vec{p}\vert}\right|
	-\frac{i\pi}{2}\theta(\vert\vec{p}\vert^2-p_0^2),
\end{equation*}
with the retarded prescription $p_0\to p_0+i\eta$, $\eta\to0^+$. In addition, $\mathcal{P}_{L,R}=(1\mp\gamma_5)/2$ are the left- and right-handed chiral projectors.
Furthermore, $m_f(T)$ is the thermal mass, which scales linearly with $T$ at leading HTL order.
The contributions of the non-Abelian $\mathrm{SU(N)}$, Abelian $\mathrm{U(1)_Y}$, and Yukawa interactions to the thermal masses are given, in order, by \cite{Bellac:2011kqa, PhysRevD.26.2789} 
\begin{equation}\label{3.5}
	m^2_f(T)=g^2\frac{T^2}{8}C_2(N) +\left(\frac{{g^\prime}Y}{2}\right)^2\frac{T^2}{8}+\lambda^2\frac{T^2}{16},
\end{equation}
where $g$ and $g^\prime$ are gauge couplings, while $\lambda$ denotes a Yukawa eigenvalue, taken real and nonnegative in the convention adopted above; $C_2(N)$ is the quadratic Casimir operator of the fermions: 
$C_2(N)=\frac{N^2-1}{2N}$ \cite{Peskin:1995ev}.
 In the collective regime, where $P\sim\xi T$, Eq.~\eqref{3.3} gives $\Sigma_{\mathrm{HTL}}\sim m_f^2/P\sim\xi T$. By contrast, the zero-temperature one-loop contribution scales as $\xi^2P$ and is therefore suppressed relative to the leading HTL term \cite{Hataei,Carignano_2018}. Its UV-divergent part is removed by the standard zero-temperature renormalization, leaving a finite contribution that enters beyond leading HTL order \cite{Hataei,Carignano_2018}.
	
To obtain the fermion thermal masses in the symmetric phase, we use the expanded form of the fermion Lagrangian in Eq.~\eqref{mass-basis-lagrangian}. For the quark sector, this yields
	\begin{align}\label{3.8}
		&\mathcal{L}^\prime_{\rm Q}=i\bar{u}^\prime_{i_R}\big[\slashed{\partial}+i\frac{2}{3}g^\prime\slashed{B}_Y\big]u^\prime_{i_R}+i\bar{d}^\prime_{i_R}\big[\slashed{\partial}-i\frac{2}{6}g^\prime\slashed{B}_Y\big]d^\prime_{i_R}\no\\
		&+i\bar{u}_{i_L}^\prime	\big[\slashed{\partial}+\frac{ig}{2}\slashed{A}_{3}+\frac{i}{6}g^\prime\slashed{B}_Y\big]u_{i_L}^\prime-\frac{g}{2}\bar{u}_{i_L}^\prime V_{ij}(\slashed{A_{1}}-i\slashed{A_{2}})d_{j_L}^\prime-\frac{g}{2}\bar{d}_{i_L}^\prime	V_{ji}^*(\slashed{A_{1}}+i\slashed{A_{2}})u_{j_L}^\prime+i\bar{d}_{i_L}^\prime\big[\slashed{\partial}-\frac{ig}{2}\slashed{A}_{3}+\frac{i}{6}g^\prime\slashed{B}_Y\big]d_{i_L}^\prime\no\\
		&-\left\{\bar{u}^\prime_{i_L}V_{ij}\lambda_{d_j}\phi^{(+)}d_{j_R}^\prime+\bar{d}^\prime_{i_L}\lambda_{d_i}\phi^{(0)}d_{i_R}^\prime+\bar{u}^\prime_{i_L}\lambda_{u_i}\phi^{(0)*}u_{i_R}^\prime-\bar{d}^\prime_{i_L}V_{ji}^*\lambda_{u_j}\phi^{(+)*}u_{j_R}^\prime\right\}-\mathrm{H.C.}.
	\end{align}
The indices \lq{\textit{i}}\rq\ and \lq{\textit{j}}\rq\ are flavor summation indices. We can also define $A^\pm_\mu=(A_{1\mu}\mp iA_{2\mu})/\sqrt{2}$. The lepton-sector Lagrangian is obtained similarly by the corresponding replacements of fields and mixing matrices. The Yukawa and charged-gauge contributions to the left-handed up-quark self-energy are shown in Fig.~\ref{left-handed1}.
	\begin{figure}
		\centering
		\begin{subfigure}{0.333\textwidth}
			\centering
			\begin{tikzpicture}
				\begin{feynman}
					\vertex (a) {};
					\vertex [right=0.82cm of a] (b) [label=below:$V^*_{kj'}\lambda_{d_{j'}}$];
					\vertex [right=3.40cm of b] (c) [label=below:$V_{ij}\lambda_{d_j}$];
					\vertex [right=0.82cm of c] (d) {};
					\vertex [right=0.72cm of b, yshift=0.38cm] {\(d^\prime_{j'_R}\)};
					\vertex [left=0.72cm of c, yshift=0.38cm] {\(d^\prime_{j_R}\)};
					\vertex [right=0.72cm of b] {\(/\)};
					\vertex [left=0.72cm of c] {\(/\)};
					\diagram*{
						(a) -- [fermion, edge label=$u^\prime_{k_L}$] (b) -- [fermion] (c) -- [fermion, edge label=$u^\prime_{i_L}$] (d),
						(b) -- [scalar, quarter left, edge label=\(\phi^{(\pm)}\), ->] (c),
					};
				\end{feynman}
			\end{tikzpicture}
			\caption{}\label{phi+exchange}
		\end{subfigure}\hfill
		\begin{subfigure}{0.333\textwidth}
			\centering
			\begin{tikzpicture}[baseline=-0.70cm]
				\begin{feynman}
					\vertex (a) {};
					\vertex [right=0.82cm of a] (b) [label=below:$\lambda_{u_k}\delta_{kj'}$];
					\vertex [right=3.40cm of b] (c) [label=below:$\lambda_{u_i}\delta_{ij}$];
					\vertex [right=0.82cm of c] (d) {};
					\vertex [right=0.72cm of b, yshift=0.38cm] {\(u^\prime_{j'_R}\)};
					\vertex [left=0.72cm of c, yshift=0.38cm] {\(u^\prime_{j_R}\)};
					\vertex [right=0.72cm of b] {\(/\)};
					\vertex [left=0.72cm of c] {\(/\)};
					\diagram*{
						(a) -- [fermion, edge label=$u^\prime_{k_L}$] (b) -- [fermion] (c) -- [fermion, edge label=$u^\prime_{i_L}$] (d),
						(b) -- [scalar, quarter left, edge label=\(\phi^{(0)}\), ->] (c),
					};
				\end{feynman}
			\end{tikzpicture}
			\caption{}\label{phi0exchange}
		\end{subfigure}\hfill
		\begin{subfigure}{0.333\textwidth}
			\centering
			\begin{tikzpicture}
				\begin{feynman}
					\vertex (a) {};
					\vertex [right=0.82cm of a] (b) [label=below:$\frac{g}{\sqrt{2}}V^*_{kj'}\gamma^\mu$];
					\vertex [right=3.40cm of b] (c) [label=below:$\frac{g}{\sqrt{2}}V_{ij}\gamma^\nu$];
					\vertex [right=0.82cm of c] (d) {};
					\vertex [right=0.72cm of b, yshift=0.38cm] {\(d^\prime_{j'_L}\)};
					\vertex [left=0.72cm of c, yshift=0.38cm] {\(d^\prime_{j_L}\)};
					\vertex [right=0.72cm of b] {\(/\)};
					\vertex [left=0.72cm of c] {\(/\)};
					\diagram*{
						(a) -- [fermion, edge label=$u^\prime_{k_L}$] (b) -- [fermion] (c) -- [fermion, edge label=$u^\prime_{i_L}$] (d),
						(b) -- [boson, quarter left, edge label=\(A^\pm\), ->] (c),
					};
				\end{feynman}
			\end{tikzpicture}
			\caption{}\label{Apm-exchange}
		\end{subfigure}
		\caption{Yukawa and charged-gauge contributions to the left-handed up-quark self-energy, with vertex factors. Incoming flavor $k$ and outgoing flavor $i$ correspond to the $ik$ matrix entry.}
		\label{left-handed1}
	\end{figure}
The self-energy corresponding to Fig.~\ref{phi+exchange} has the schematic form
	\begin{equation}\label{4.2}
		\begin{split}
		\Sigma^{(a)}_{ik,L}\sim-\big[\delta_{jj^\prime} V^*_{kj'}\lambda_{d_{j'}} V_{ij}\lambda_{d_j}\big]\, T\sum_{n}\int\frac{d^3k}{(2\pi)^3}\bigl[\textrm{Loop contributions}\bigr]\mathcal{P}_L.\\
		\end{split}
	\end{equation}
	In this case, the thermal mass becomes
	\begin{align}
		(m^2_{u^\prime_{{ik}_L}})^{\phi^{(\pm)}}=\frac{V_{ij}\lambda_{d_j}^2 V^*_{kj}}{16}T^2,
	\end{align}
	indicating that the left-handed self-energy induced by virtual $\phi^{(\pm)}$ exchange is flavor non-diagonal. In contrast, the contributions corresponding to Figs.~\ref{phi0exchange}--\ref{Apm-exchange} give
	\begin{align}
		(m^2_{u'_{{ik}_L}})^{\phi^{(0)}}=\frac{\lambda_{u_i}^2T^2}{16}\delta_{ik},\quad 	(m^2_{u'_{{ik}_L}})^{A^\pm}=\frac{g^2T^2}{16}\delta_{ik},
	\end{align}
	and are flavor diagonal. The same reasoning applies to the remaining sectors: for left-handed fermions, only virtual $\phi^{(\pm)}$ exchange generates flavor non-diagonal thermal contributions. For right-handed fermions, all contributions remain flavor diagonal, including those involving virtual $\phi^{(\pm)}$ exchange. For example, for a right-handed up quark,
	\begin{align}
		(m^2_{u'_{i_R}})^{\phi^{(\pm)}}=\frac{V^*_{ij}\lambda_{u_i}^2V_{ij}T^2}{16}=\frac{\lambda_{u_i}^2T^2}{16},
	\end{align}
	where the unitarity of the CKM matrix has been used.
	The remaining cases follow in the same way, and the final results are\footnote{We compute the thermal masses perturbatively, including the top-Yukawa contribution, following Refs.~\cite{PhysRevD.47.3546,Katz_2014,PhysRevD.45.2933}.}:
				\begin{align}\label{3.9}
						m^2_{u'_{i_R}}
					&=\frac{{g^\prime}^2T^2}{18}
					+\frac{{g^{\prime\prime}}^2T^2}{6}
					+\frac{\lambda_{u_i}^2T^2}{8},
					&\qquad
					m^2_{d'_{i_R}}
					&=\frac{{g^\prime}^2T^2}{72}
					+\frac{{g^{\prime\prime}}^2T^2}{6}
					+\frac{\lambda_{d_i}^2T^2}{8},\no
					\\
					m^2_{\nu'_{i_R}}
					&=\frac{\lambda_{\nu_i}^2T^2}{8},
					&
					m^2_{e'_{i_R}}
					&=\frac{{g^\prime}^2T^2}{8}
					+\frac{\lambda_{e_i}^2T^2}{8},\no
					\\
					m^2_{u'_{{ik}_L}}
					&=G_qT^2\delta_{ik}
					+\left(
					\lambda_{u_i}^2\delta_{ik}
					+V_{ij}\lambda_{d_j}^2V^*_{kj}
					\right)\frac{T^2}{16},
					&\qquad
					m^2_{d'_{{ik}_L}}
					&=G_qT^2\delta_{ik}
					+\left(
					\lambda_{d_i}^2\delta_{ik}
					+V^*_{ji}\lambda_{u_j}^2V_{jk}
					\right)\frac{T^2}{16},\no
					\\
					m^2_{\nu'_{{ik}_L}}
					&=G_\ell T^2\delta_{ik}
					+\left(
					\lambda_{\nu_i}^2\delta_{ik}
					+U_{ij}\lambda_{e_j}^2U^*_{kj}
					\right)\frac{T^2}{16},
					&\qquad
					m^2_{e'_{{ik}_L}}
					&=G_\ell T^2\delta_{ik}
					+\left(
					\lambda_{e_i}^2\delta_{ik}
					+U^*_{ji}\lambda_{\nu_j}^2U_{jk}
					\right)\frac{T^2}{16},
				\end{align}
	with the gauge coefficients
	\begin{equation}\label{eq:Gdefs}
		G_q=\frac{{g^\prime}^2}{288}+\frac{3g^2}{32}+\frac{{g^{\prime\prime}}^2}{6},
		\qquad
		G_\ell=\frac{{g^\prime}^2}{32}+\frac{3g^2}{32}.
	\end{equation}
Repeated generation indices are summed over. 
	We have also included the $\mathrm{SU}(3)_C$ contribution, with strong-interaction  coupling $g^{\prime\prime}$, using Eq.~\eqref{3.5}.
	In the SM limit, the corresponding lepton thermal masses reduce to
	\begin{equation}\label{3.12}
		\begin{split}
			&m^2_{{i}_{L^\prime}}=\frac{{g^\prime}^2T^2}{32}+\frac{3g^2T^2}{32}+\frac{\lambda_{e_i}^2T^2}{16},\quad	m^2_{e^\prime_{i_{R  }}}=\frac{{g^\prime}^2T^2}{8}+\frac{\lambda_{e_i}^2T^2}{8}.
		\end{split}
	\end{equation} 
	Here, the index \lq{$\textit{L}^\prime$}\rq\ refers to both the left-handed neutrinos and charged leptons. In this case, the Yukawa contribution to the right-handed charged-lepton thermal mass is twice that of the left-handed lepton doublet, reflecting the different Higgs channels available to the two chiralities. Since the SM lepton-sector Lagrangian is flavor diagonal, no flavor mixing is induced in the corresponding self-energies.
	Thermal masses and flavor effects for SM fermions were obtained in Ref.~\cite{Gavela_1994}. Thermal masses were also derived in the baryogenesis context in Ref.~\cite{Davidson_1994}, where CKM effects were neglected.
	
For clarity, the left-handed up-quark thermal mass matrix is explicitly
\begin{equation}\label{4.9}
\mathcal M_{u'_L}^2(T)=G_qT^2\mathbbm{1}+\frac{T^2}{16}
\begin{pmatrix}
\lambda_{u_1}^2+|V_{1j}|^2\lambda_{d_j}^2 & V_{1j}\lambda_{d_j}^2V_{2j}^*& V_{1j}\lambda_{d_j}^2V_{3j}^*\\
 V_{2j}\lambda_{d_j}^2V_{1j}^*&\lambda_{u_2}^2+|V_{2j}|^2\lambda_{d_j}^2& V_{2j}\lambda_{d_j}^2V_{3j}^*\\
 V_{3j}\lambda_{d_j}^2V_{1j}^*& V_{3j}\lambda_{d_j}^2V_{2j}^*&\lambda_{u_3}^2+|V_{3j}|^2\lambda_{d_j}^2
\end{pmatrix},
\end{equation}
The off-diagonal entries, which are in general nonzero, show that the broken-phase mass basis does not diagonalize the leading HTL propagator in the symmetric phase. The momentum dependence is universal across flavors, Eq.~\eqref{3.3}, which differ only through their thermal mass matrices. Therefore, diagonalizing these matrices is sufficient to identify the fermionic quasiparticle modes propagating in the symmetric phase at leading HTL order.

\section{Symmetric-Phase Thermal Basis}\label{sec.4}
	We now diagonalize the left-handed thermal matrices and rewrite the theory in the resulting symmetric-phase thermal basis.
 The left-handed entries of Eq.~\eqref{3.9} take the matrix form
	\begin{align}\label{eq:massbasis-matrices}
		&\mathcal M^2_{u^\prime_L}(T)=T^2\Big[G_q\mathbbm{1}+\tfrac{1}{16}H_u\Big],
		&&H_u\equiv \lambda_u^2+V\lambda_d^2V^\dagger,\nonumber\\
		&\mathcal M^2_{d^\prime_L}(T)=T^2\Big[G_q\mathbbm{1}+\tfrac{1}{16}H_d\Big],
		&&H_d\equiv \lambda_d^2+V^\dagger \lambda_u^2V=V^\dagger H_uV,\nonumber\\
		&\mathcal M^2_{\nu^\prime_L}(T)=T^2\Big[G_\ell\mathbbm{1}+\tfrac{1}{16}H_\nu\Big],
		&&H_\nu\equiv \lambda_\nu^2+U\lambda_e^2U^\dagger,\nonumber\\
		&\mathcal M^2_{e^\prime_L}(T)=T^2\Big[G_\ell\mathbbm{1}+\tfrac{1}{16}H_e\Big],
		&&H_e\equiv \lambda_e^2+U^\dagger \lambda_\nu^2U=U^\dagger H_\nu U,
	\end{align}
	while the right-handed matrices in Eq.~\eqref{3.9} are already diagonal.
	Since $H_u$ and $H_\nu$ are Hermitian, they admit orthonormal eigenvectors.
	Let $r_{q_a}$ and $r_{\ell_a}$ denote these eigenvectors,
	\begin{equation}
		H_u r_{q_a}=h_{q_a}r_{q_a},
		\qquad
		H_\nu r_{\ell_a}=h_{\ell_a}r_{\ell_a},
	\end{equation}
	and collect them as the columns of the unitary matrices
	$R_q=(r_{q_1},r_{q_2},r_{q_3})$ and
	$R_\ell=(r_{\ell_1},r_{\ell_2},r_{\ell_3})$. Then
	\begin{equation}\label{eq:Rdefs}
		R_q^\dagger H_uR_q
		=H_q^{(T)}
		\equiv\mathrm{diag}(h_{q_1},h_{q_2},h_{q_3}),
		\qquad
		R_\ell^\dagger H_\nu R_\ell
		=H_\ell^{(T)}
		\equiv\mathrm{diag}(h_{\ell_1},h_{\ell_2},h_{\ell_3}).
	\end{equation}
	The unitary equivalence $H_dV^\dagger=V^\dagger H_u$ then determines the
	corresponding eigenvectors of $H_d$:
	\begin{equation}
		(H_dV^\dagger)r_{q_a}
		=(V^\dagger H_u)r_{q_a}
		=h_{q_a}(V^\dagger r_{q_a}),
	\end{equation}
	so $V^\dagger r_{q_a}$ is an eigenvector of $H_d$ with the same
	eigenvalue $h_{q_a}$. The same argument applies to the lepton sector. It follows that
	\begin{equation}
		(V^\dagger R_q)^\dagger H_d(V^\dagger R_q)
		=H_q^{(T)},
		\qquad
		(U^\dagger R_\ell)^\dagger H_e(U^\dagger R_\ell)
		=H_\ell^{(T)}.
	\end{equation}
	Thus, the two members of each left-handed weak doublet have identical
	thermal eigenvalues, while their eigenvectors are related by the corresponding CKM or PMNS transformation. 
	
	In analogy with the rotations of Sec.~\ref{sec. 2}, we define the symmetric-phase thermal basis by rotating only the left-handed weak doublets,
	\begin{align}\label{eq:thermal-rotations}
		&Q^\prime\rightarrow\mathcal R_Q\,Q^{(T)}
		=\mathrm{diag}\big(R_q,\,V^\dagger R_q\big)\,Q^{(T)},\nonumber\\
		&L^\prime\rightarrow\mathcal R_L\,L^{(T)}
		=\mathrm{diag}\big(R_\ell,\,U^\dagger R_\ell\big)\,L^{(T)}.
	\end{align}
	Here $Q^{(T)}=\mathrm{Col}\big(U_L,D_L\big)$ and
	$L^{(T)}=\mathrm{Col}\big(N_L,E_L\big)$. No transformation is applied to the right-handed singlets $u_R^\prime$, $d_R^\prime$, $\nu_R^\prime$, and $e_R^\prime$: they remain exactly the fields of the broken-phase mass-basis convention introduced in Sec.~\ref{sec. 2}. This is permitted because, at $\langle\Phi\rangle=0$, no Dirac mass term ties the two chiralities together. 
	
	Applying Eq.~\eqref{eq:thermal-rotations} to the broken-phase mass-basis Lagrangian,
	Eq.~\eqref{mass-basis-lagrangian}, we proceed exactly as in
	Sec.~\ref{sec. 2}. For the left-handed gauge sector,
	\begin{equation}\label{eq:thermal-gauge}
		i\bar Q^{(T)}
		\big(\mathcal R_Q^\dagger U_Q^\dagger\slashed{D}U_Q\mathcal R_Q\big)
		Q^{(T)}
		=i\bar Q^{(T)}\slashed{D}Q^{(T)},
	\end{equation}
	since
	$R_q^\dagger V\big(V^\dagger R_q\big)=\mathbbm{1}$
	in the off-diagonal blocks.
	For the Yukawa sector, the matrices of Eq.~\eqref{quark-sector-mass}
	transform as
	\begin{equation}\label{eq:V-collapse}
		\mathcal R_Q^\dagger\,\mathcal V^\prime\lambda_d
		=
		\begin{pmatrix}
			R_q^\dagger&0\\
			0&R_q^\dagger V
		\end{pmatrix}
		\begin{pmatrix}
			V\lambda_d&0\\
			0&\lambda_d
		\end{pmatrix}
		=\widetilde Y_d\,\mathbbm{1},
		\qquad
		\mathcal R_Q^\dagger\,\mathcal V^{\prime\prime}\lambda_u
		=\widetilde Y_u
		\mathbbm{1},
	\end{equation}
	with the thermal Yukawa matrices defined as
	\begin{equation}\label{eq:tildeY-defs}
		\widetilde Y_u\equiv R_q^\dagger \lambda_u,
		\qquad
		\widetilde Y_d\equiv R_q^\dagger V\lambda_d,
		\qquad
		\widetilde Y_\nu\equiv R_\ell^\dagger \lambda_\nu,
		\qquad
		\widetilde Y_e\equiv R_\ell^\dagger U\lambda_e.
	\end{equation}
	The fermionic Lagrangian after this left-handed rotation is therefore
	\begin{subequations}\label{eq:thermal-lagrangian}
		\begin{align}
			\mathcal L^{(T)}_Q
			&=i\bar Q^{(T)}\slashed{D}Q^{(T)}
			+i\bar u_R^\prime\slashed{D}u_R^\prime
			+i\bar d_R^\prime\slashed{D}d_R^\prime
			-\Big\{
			\bar Q^{(T)}\widetilde Y_d\,\Phi\,d_R^\prime
			+\bar Q^{(T)}\widetilde Y_u\,\tilde\Phi\,u_R^\prime
			\Big\}
			-\mathrm{H.C.},
			\label{eq:thermal-lagrangian-q}\\
			\mathcal L^{(T)}_L
			&=i\bar L^{(T)}\slashed{D}L^{(T)}
			+i\bar e_R^\prime\slashed{D}e_R^\prime
			+i\bar\nu_R^\prime\slashed{\partial}\nu_R^\prime
			-\Big\{
			\bar L^{(T)}\widetilde Y_e\,\Phi\,e_R^\prime
			+\bar L^{(T)}\widetilde Y_\nu\,\tilde\Phi\,\nu_R^\prime
			\Big\}
			-\mathrm{H.C.}.
			\label{eq:thermal-lagrangian-l}
		\end{align}
	\end{subequations}
Thus, the gauge interactions remain diagonal in the symmetric-phase thermal basis, whereas the transformed Yukawa couplings are generally non-diagonal. 
	\subsection{HTL spectrum and propagators}\label{sec.4C}
	The one-loop HTL calculations of Sec.~\ref{sec. 3} can be directly applied
	in the symmetric-phase thermal basis. For the left-handed up-type fields, for example $U_L$,
	the neutral- and charged-scalar exchanges contribute the flavor structures
	$\widetilde Y_u\widetilde Y_u^\dagger$ and
	$\widetilde Y_d\widetilde Y_d^\dagger$, respectively. For $D_L$, the same two
	structures appear with the scalar channels interchanged. For example, the
	charged- and neutral-scalar contributions to the $U_L$ thermal mass matrix are
	\begin{equation}
		\left(m^2_{U_{ab,L}}\right)^{\phi^{(\pm)}}
		+
		\left(m^2_{U_{ab,L}}\right)^{\phi^{(0)}}
		=
		\left[
		\widetilde Y_d\widetilde Y_d^\dagger+
		\widetilde Y_u\widetilde Y_u^\dagger
		\right]_{ab}
		\frac{T^2}{16}
		=
		(R_q^\dagger)_{ai}
		\left[
		\lambda_{u_i}^2\delta_{im}
		+
		V_{ij}\lambda_{d_j}^2(V^\dagger)_{jm}
		\right]
		(R_q)_{mb}
		\frac{T^2}{16}
		=
		(H_q^{(T)})_{ab}\frac{T^2}{16}.
	\end{equation}
	Thus, although $\widetilde Y_u$ and $\widetilde Y_d$ are individually
	non-diagonal, their left-handed combination is diagonal by construction.
	
	For the right-handed up-type quark fields, the scalar contributions instead
	involve the combination
	\begin{align}
		\left(m^2_{u^\prime_{ik,R}}\right)^{\phi^{(\pm)}}
		+
		\left(m^2_{u^\prime_{ik,R}}\right)^{\phi^{(0)}}
		&=
		2(\widetilde Y_u^\dagger\widetilde Y_u)_{ik}
		\frac{T^2}{16}=
		2\bigl[\lambda_u R_qR_q^\dagger\lambda_u\bigr]_{ik}
		\frac{T^2}{16}
		=
		\frac{\lambda_{u_i}^2T^2}{8}\delta_{ik},
	\end{align}
	where the unitarity of $R_q$ has been used. The same applies to the other
	right-handed fields. 
	The new thermal eigenvalues introduced in this section are therefore only the
	left-handed ones,
	\begin{equation}\label{eq:thermal-masses-final}
		m^2_{U_L,a}(T)=m^2_{D_L,a}(T)
		=
		T^2\left(G_q+\frac{h_{q_a}}{16}\right),
		\qquad
		m^2_{N_L,a}(T)=m^2_{E_L,a}(T)
		=
		T^2\left(G_\ell+\frac{h_{\ell_a}}{16}\right).
	\end{equation}

	Now, let \(H\) denote \(H_u\) in the quark sector or \(H_\nu\) in the lepton sector. Since \(H\) is Hermitian,
	its three eigenvalues \(h_a\) are real and satisfy 
	\begin{align}
	(H-h_a\mathbbm{1})r_a=0.
	\end{align}
	The eigenvalues satisfy
	\begin{equation}
		\det\!\left(h\mathbbm{1}-H\right)
		=
		h^3-I_1h^2+I_2h-I_3=0 .
	\end{equation}
	where
	\begin{equation}\label{eq:invariants}
		I_1=\tr H,
		\qquad
		I_2=\frac{1}{2}\left[(\tr H)^2-\tr(H^2)\right],
		\qquad
		I_3=\det H .
	\end{equation}
	The characteristic equation has three real roots, which can be written using the trigonometric cubic solution (see Sec.~5.6 of Ref.~\cite{Press:2007ipz}).
	Defining
	\begin{equation}
		\phi \equiv
		\frac{1}{3}\arccos\!\left(
		\frac{2I_1^3-9I_1I_2+27I_3}
		{2\left(I_1^2-3I_2\right)^{3/2}}
		\right),
		\qquad
		0\leq\phi\leq\frac{\pi}{3},
	\end{equation}
	the three eigenvalues, ordered as $h_1< h_2< h_3$, are
	\begin{align}
		h_1 &=
		\frac{I_1}{3}
		+\frac{2}{3}\sqrt{I_1^2-3I_2}\,
		\cos\!\left(\phi+\frac{2\pi}{3}\right),
		\nonumber\\
		h_2 &=
		\frac{I_1}{3}
		+\frac{2}{3}\sqrt{I_1^2-3I_2}\,
		\cos\!\left(\phi-\frac{2\pi}{3}\right),
		\nonumber\\
		h_3 &=
		\frac{I_1}{3}
		+\frac{2}{3}\sqrt{I_1^2-3I_2}\,
		\cos\phi .
		\label{eq:trig-solution}
	\end{align}
	For the quark sector, the three invariants entering the characteristic
	polynomial are explicitly
	\begin{align}\label{eq:invariants-explicit}
		&I_{q,1}
		=
		\sum_i \lambda_{u_i}^2
		+\sum_j \lambda_{d_j}^2 ,
		\\
		&I_{q,2}=
		\frac{1}{2}
		\Bigl[
		I_{q,1}^2
		-\sum_i \lambda_{u_i}^4
		-\sum_j \lambda_{d_j}^4
		-2\sum_{i,j}
		\lambda_{u_i}^2\lambda_{d_j}^2
		|V_{ij}|^2\Bigr],
		\\
		&I_{q,3}
		=
		\det\!\left[
		\lambda_u^2
		+V\lambda_d^2V^\dagger
		\right].
	\end{align}
	The lepton-sector expressions follow from the replacements
	\[
	(u,d,V)\rightarrow(\nu,e,U).
	\]
The invariants $I_1$, $I_2$, and $I_3$ are unchanged under arbitrary unitary weak-basis transformations of the left-handed doublets. Moreover, CP conjugation changes $H$ to $H^*$ without changing its eigenvalues, so these spectra are CP even. By contrast, the
	components of the corresponding eigenvectors depend on the phase
	convention chosen for the flavor basis.
	
	In the symmetric-phase thermal basis, each eigenmode obeys the standard HTL
	dispersion relation independently \cite{Bellac:2011kqa}. For each left- or right-handed fermionic mode, the corresponding inverse propagator takes the form
	\begin{align}
		S_a^{-1}(P)=\gamma^0A_a(P)-\boldsymbol{\gamma}\cdot\hat{\mathbf p}\,B_a(P)
	\end{align}
	with
	\begin{equation}\label{eq:AB}
		A_a(P)=p_0-\frac{m_a^2(T)}{p}\,\mathcal Q_0\!\left(\frac{p_0}{p}\right),
		\qquad
		B_a(P)=p+\frac{m_a^2(T)}{p}\left[1-\frac{p_0}{p}\,\mathcal Q_0\!\left(\frac{p_0}{p}\right)\right],
	\end{equation}
	the dispersion relations $A_a(P)=\pm B_a(P)$ yield, for every mode, the usual normal and plasmino branches \cite{Bellac:2011kqa} with a mode-dependent coefficient. Although all fermions are massless at tree level in the symmetric phase, Tab.~\ref{tab:thermal-numerics} exhibits distinct leading thermal eigenvalues for the generation-space modes; establishing whether their spectral peaks are resolved requires a damping-rate analysis.
	\subsection{HTL Effective Lagrangian}
The fermionic part of the leading HTL effective action \cite{PhysRevD.45.R1827,BRAATEN1990310,Taylor:1990ia,FRENKEL1992156}
can be expressed in the symmetric-phase thermal basis. Its propagators and two-fermion--$N$-gauge-boson HTL vertices are diagonal in generation labels. Taking $v^\mu=(1,\hat{\mathbf v})$, one has
	\begin{align}\label{eq:HTL-diagonal}
		\mathcal L^{(T)}_{\mathrm{HTL}}
		=&-\bar Q^{(T)}\,\mathcal M^2_{Q_L}(T)\int\frac{d\Omega_{\mathbf v}}{4\pi}\,\frac{\slashed{v}}{i v\cdot D^{(T)}_Q}\,Q^{(T)}
		-\bar L^{(T)}\,\mathcal M^2_{L_L}(T)\int\frac{d\Omega_{\mathbf v}}{4\pi}\,\frac{\slashed{v}}{i v\cdot D^{(T)}_L}\,L^{(T)}\nonumber\\
		&-\sum_{\chi=u,d,e,\nu}\bar\chi_R^\prime\,\mathcal M^2_{\chi_R^\prime}(T)\int\frac{d\Omega_{\mathbf v}}{4\pi}\,\frac{\slashed{v}}{i v\cdot D_{\chi_R^\prime}}\,\chi_R^\prime,
	\end{align}
The left-handed mass matrices are diagonal by Eqs.~\eqref{eq:Rdefs} and \eqref{eq:thermal-masses-final}, as are the right-handed matrices in Eq.~\eqref{3.9}. The corresponding covariant derivatives carry no thermal-mode structure, as shown in Eq.~\eqref{eq:thermal-gauge}.
	Expanding $(v\cdot D^{(T)})^{-1}$ generates higher-point functions and preserves the same diagonal thermal-mode structure for all HTL gauge
	two-fermion--$N$-gauge-boson vertices.
	
The diagonal fermionic HTL effective action also simplifies higher-order calculations. For soft fermions, the refined gauge-theory power counting of Ref.~\cite{MirzaCarrington:2013} identifies one-soft-loop diagrams with HTL-resummed propagators and vertices as the relevant NLO gauge contribution . Since these gauge vertices preserve thermal-mode labels, the corresponding  self-energy has the schematic structure
\begin{equation}\label{eq:NLO-gauge}
\bigl(\delta\Sigma_{\mathrm{NLO}}^{R,g}\bigr)_{ab}(P)
=\delta_{ab}\,\delta\Sigma_{a}^{R,g}(P),\qquad
\delta\Sigma_{a}^{R,g}\sim \xi^2T,
\end{equation}
where $\xi$ denotes the gauge-coupling power-counting parameter. The imaginary parts of these retarded corrections determine the gauge-induced damping rates after the appropriate pole projection. The complete electroweak NLO self-energy can also receive Yukawa contributions, whose flavor structures need not be diagonal in the leading symmetric-phase thermal basis. The required bosonic HTL propagators and matching contributions lie outside the fermionic action displayed here.
	\subsection{Numerical illustration}\label{sec.4E}
For illustration, we use representative electroweak-scale Yukawa inputs, without evolving the couplings to the plasma temperature, as follows~\cite{ParticleDataGroup:2024cfk,ParticleDataGroup:2026}
	\begin{align*}
		&\lambda_{u_1}=1.3\times10^{-5},\qquad
		\lambda_{u_2}=7.3\times10^{-3},\qquad
		\lambda_{u_3}=0.995,\\
		&\lambda_{d_1}=2.7\times10^{-5},\qquad
		\lambda_{d_2}=5.5\times10^{-4},\qquad
		\lambda_{d_3}=2.4\times10^{-2},\\
		&\lambda_{e_1}\simeq2.94\times10^{-6},\qquad
		\lambda_{e_2}\simeq6.07\times10^{-4},\qquad
		\lambda_{e_3}\simeq1.02\times10^{-2}.
	\end{align*}
	For the Dirac-neutrino Yukawa couplings, we use
	$\lambda_{\nu_i}=\sqrt{2}\,m_{\nu_i}/v$, with \(v\simeq246.22\) GeV. Taking a representative
	normal-ordering benchmark with $m_{\nu_1}\simeq0$, the measured
	mass-squared splittings from the NuFIT 6.1 normal-ordering fit~\cite{NuFIT:2025v61} give
	\[
	\lambda_{\nu_1}\simeq0,\qquad
\lambda_{\nu_2}\simeq4.99\times10^{-14},\qquad
\lambda_{\nu_3}\simeq2.88\times10^{-13}\BlueEdit{.}
	\]
	Their contributions to the numerical thermal coefficients are
	therefore negligible at the precision displayed below.
Using the CKM and PMNS inputs presented in App.~\ref{mixing-matrices},
	Tab.~\ref{tab:thermal-numerics} compares the diagonal broken-phase mass-basis entries, corresponding to an approximation in which off-diagonal components are neglected, with the eigenvalues of the complete leading Yukawa flavor matrices. Only the Yukawa coefficients are tabulated since adding the  gauge contribution $G_{q,\ell}$ shifts all squared thermal-mass eigenvalues equally.
	\begin{table}[H]
        \caption{\label{tab:thermal-numerics}%
        Numerical comparison of the left-handed Yukawa thermal coefficients in the broken-phase mass basis and the symmetric-phase thermal basis. For quarks, the two diagonal columns are $(H_u)_{aa}/16$ and $(H_d)_{aa}/16$; for leptons, they are $(H_\nu)_{aa}/16$ and $(H_e)_{aa}/16$. The final column gives $h_a/16$.}
        \begin{ruledtabular}
            \begin{tabular}{ccccc}
                Sector
                & $a$
                & \shortstack{Broken-phase\\mass basis\\$(H_{u,\nu})_{aa}/16$}
                & \shortstack{Broken-phase\\mass basis\\$(H_{d,e})_{aa}/16$}
                & \shortstack{Symmetric-phase\\thermal basis\\$h_a/16$}
                \\
                \colrule
                Quarks
				& $1$
				& $1.52214259\times10^{-9}$
				& $4.77311338\times10^{-6}$
				& $1.50183701\times10^{-9}$
				\\
				& $2$
				& $3.41171386\times10^{-6}$
				& $1.08024540\times10^{-4}$
				& $3.41169724\times10^{-6}$
				\\
				& $3$
				& $6.19124989\times10^{-2}$
				& $6.18031144\times10^{-2}$
				& $6.19124989\times10^{-2}$
				\\
				\colrule
				Leptons
				& $1$
				& $1.38117437\times10^{-6}$
				& $5.40225000\times10^{-13}$
				& $5.40225000\times10^{-13}$
				\\
				& $2$
				& $1.76492272\times10^{-6}$
				& $2.30280625\times10^{-8}$
				& $2.30280625\times10^{-8}$
				\\
				& $3$
				& $3.37943151\times10^{-6}$
				& $6.50250000\times10^{-6}$
				& $6.50250000\times10^{-6}$
				\\
			\end{tabular}
        \end{ruledtabular}
    \end{table}
	
	\section{Summary and Conclusion}
\label{sec.conclusion}
Within the minimally extended Standard Model with Dirac neutrinos, we have
determined the symmetric-phase thermal basis in which the leading HTL self-energies and the corresponding effective action are diagonal in the electroweak symmetric phase.
Charged-scalar Yukawa interactions make the left-handed thermal matrices
generally non-diagonal in the {broken-phase mass basis, whereas the
right-handed thermal matrices remain diagonal. Since the HTL momentum
dependence factorizes from the flavor matrices, the propagating modes can be
identified by momentum-independent unitary rotations. That is, although the tree-level masses are zero, the plasma identifies the distinct quasiparticles through their interactions. The matrices of the
two members of each left-handed weak doublet are unitarily equivalent and
therefore possess identical leading-HTL thermal spectra. A common
weak-doublet rotation diagonalizes their propagators while preserving
diagonal electroweak gauge currents; the transformed Yukawa vertices remain
generally non-diagonal, although their combined leading-HTL self-energy is
diagonal.

The thermal eigenvalues are obtained analytically from basis-independent
invariants and are CP-even. Our numerical benchmark shows that diagonal
Yukawa thermal coefficients in the separate broken-phase mass bases
need not coincide with the eigenvalues of the complete Yukawa flavor
matrices. The universal gauge contribution shifts all squared thermal
masses equally within each sector, without changing their eigenvectors.
Thus, the leading-HTL symmetric-phase thermal basis is generally distinct from the
broken-phase mass basis. Within leading HTL accuracy,
this construction yields a mode-diagonal fermionic HTL effective action
and provides a consistent starting point for flavor-sensitive thermal
calculations. Higher-order Yukawa corrections may mix these leading thermal eigenmodes, so the same fixed basis is not asserted to diagonalize the full self-energy to all orders.
	\begin{appendices}
		\begin{appendix}
			\section{CKM and PMNS Matrices}\label{mixing-matrices}
For the numerical illustration, we use the CKM moduli obtained from the 2026 Particle Data Group global fit~\cite{ParticleDataGroup:2026} and the PMNS moduli from the NuFIT 6.1 normal-ordering fit~\cite{Esteban:2024eli,NuFIT:2025v61}:
\begin{subequations}\label{eq:mixing-matrix-moduli}
\begin{align}
|V|&\simeq
\begin{pmatrix}
0.97431259 & 0.22516841 & 0.00376300\\
0.22503457 & 0.97344989 & 0.04188970\\
0.00862627 & 0.04116424 & 0.99911515
\end{pmatrix},\\
|U_{\rm PMNS}|&\simeq
\begin{pmatrix}
0.82198651 & 0.54941621 & 0.14993332\\
0.33515574 & 0.65439761 & 0.67781590\\
0.46044414 & 0.51952439 & 0.71978163
\end{pmatrix}.
\end{align}
\end{subequations}
In our convention, $U=U_{\rm PMNS}^\dagger$, and hence $|U|=|U_{\rm PMNS}|^T$.
The thermal eigenvalues and the entries in Table~\ref{tab:thermal-numerics} depend only on the squared moduli. 
		\end{appendix}
	\end{appendices}
	\bibliographystyle{apsrev4-2}
\bibliography{Refs}
\end{document}